\documentclass[sigplan,screen]{acmart}

\setcopyright{rightsretained}
\acmPrice{}
\acmDOI{10.1145/3623504.3623566}
\acmYear{2023}
\copyrightyear{2023}
\acmSubmissionID{splashws23paintmain-id7-p}
\acmISBN{979-8-4007-0399-7/23/10}
\acmConference[PAINT '23]{Proceedings of the 2nd ACM SIGPLAN International Workshop on Programming Abstractions and Interactive Notations, Tools, and Environments}{October 23, 2023}{Cascais, Portugal}
\acmBooktitle{Proceedings of the 2nd ACM SIGPLAN International Workshop on Programming Abstractions and Interactive Notations, Tools, and Environments (PAINT '23), October 23, 2023, Cascais, Portugal}
\received{2023-07-17}
\received[accepted]{2023-08-07}

\usepackage{booktabs}
\usepackage{balance}
\usepackage{listings}
\usepackage{xcolor}
\usepackage{microtype}

\definecolor{codegreen}{rgb}{0,0.6,0}
\definecolor{codegray}{rgb}{0.5,0.5,0.5}
\definecolor{codepurple}{rgb}{0.58,0,0.82}
\definecolor{backcolour}{rgb}{0.95,0.95,0.92}
\lstdefinestyle{codestyle}{
    backgroundcolor=\color{backcolour},   
    commentstyle=\color{codegreen},
    keywordstyle=\color{magenta},
    numberstyle=\tiny\color{codegray},
    stringstyle=\color{codepurple},
    basicstyle=\ttfamily\footnotesize,
    breakatwhitespace=false,         
    breaklines=true,                 
    captionpos=b,                    
    keepspaces=true,              
    numbers=left,                    
    numbersep=4pt,                  
    showspaces=false,                
    showstringspaces=false,
    showtabs=false,                  
    tabsize=2,
}

\lstdefinelanguage{Wing}{
    morekeywords={bring, let, new, inflight
        ldr, str,
        r0, r1, r2, r3, r4, r5, r6, r7, rr8, r9, r10, r11, r12,
        inflight  },
    sensitive=false, % keywords are not case-sensitive
    morecomment=[l]{//}, % l is for line comment
    morecomment=[s]{/*}{*/}, % s is for start and end delimiter
    morestring=[b]" % defines that strings are enclosed in double quotes
} %

\begin{document}

\title{A Penny a Function: Towards Cost Transparent Cloud~Programming}

\author{Lukas Böhme}
\orcid{0000-0002-3065-6997}
\email{lukas.boehme@hpi.uni-potsdam.de}
\affiliation{%
        \department{Hasso Plattner Institute}
        \institution{University of Potsdam}
        \city{Potsdam}
        \country{Germany}
}

\author{Tom Beckmann}
\orcid{0000-0003-0015-1717}
\email{tom.beckmann@hpi.uni-potsdam.de}
\affiliation{%
        \department{Hasso Plattner Institute}
        \institution{University of Potsdam}
        \city{Potsdam}
        \country{Germany}
}

\author{Sebastian Baltes}
\orcid{0000-0002-2442-7522}
\email{sebastian.baltes@adelaide.edu.au}
\affiliation{%
  \institution{University of Adelaide}
  \city{Adelaide}
  \country{Australia}
}

\author{Robert Hirschfeld}
\orcid{0000-0002-4249-6003}
\email{robert.hirschfeld@uni-potsdam.de}
\affiliation{%
        \department{Hasso Plattner Institute}
        \institution{University of Potsdam}
        \city{Potsdam}
        \country{Germany}
}

\renewcommand{\shortauthors}{Böhme, Beckmann, Baltes, Hirschfeld}

\begin{abstract}
% provide background factual infromation
%Managing and understanding monetary cost factors is crucial when developing serverless applications.
%However, the diverse range of cost factors for computation, storage and networking associated with serverless applications poses a challenge for developers in proactively managing %and minimizing costs.
%Current tools to understand and comprehend cost factors are detached from the written code, causing opaqueness regarding the origin of costs. % What about historical data
%This paper presents initial work toward a cost model based on a directed graph to calculate and derive monetary costs directly from code.
%Based on the cost model, we explore visualizations in an code editor to display costs directly tied to code, making cost exploration an integrated part of the developer experience, thereby removing the overhead of external tooling for cost comprehension of serverless applications.

% Sebastian:
Understanding and managing monetary cost factors is crucial when developing cloud applications.
However, the diverse range of factors influencing costs for computation, storage, and networking in cloud applications poses a challenge for developers who want to manage and minimize costs proactively.
Existing tools for understanding cost factors are often detached from source code, causing opaqueness regarding the origin of costs. 
Moreover, existing cost models for cloud applications focus on specific factors such as compute resources and necessitate manual effort to create the models.
This paper presents initial work toward a cost model based on a directed graph that allows deriving monetary cost estimations directly from code using static analysis.
Leveraging the cost model, we explore visualizations embedded in a code editor that display costs close to the code causing them.
This makes cost exploration an integrated part of the developer experience, thereby removing the overhead of external tooling for cost estimation of cloud applications at development time.
% How to improve?

% 

% Notes

% provide background factual infromation
% combine the method, the general aim and the specific aim of the study in one sentence
% summarise the methodogology and provide details
% indicate the achievement of the study
% present the implications of the study

% Costs are an important measurement for programming serverless applications.
% The execution of serverless applications have many different cost facots which are detachted from the code, lying in configuration files or viewable in the developer console of cloud vendors.
% This introduces cost intransparency and disallows live exploration of cost factors of an application
% In this paper, we present an model based on a directed graph to calculate costs directly tied to the code and explore ways to visualize it
% We demonstate the fisibility and open up questions for future research.
\end{abstract}



%\received{12 July 2023}
%\received[revised]{} TODO Fill me as soon as possible
%\received[accepted]{09 August 2023}

\maketitle

\newcommand{\Wing}{{\emph Wing}} 
 
\section{Introduction}
\label{sec:intro}

\begin{figure}
    \centering
    \includegraphics[width=\columnwidth]{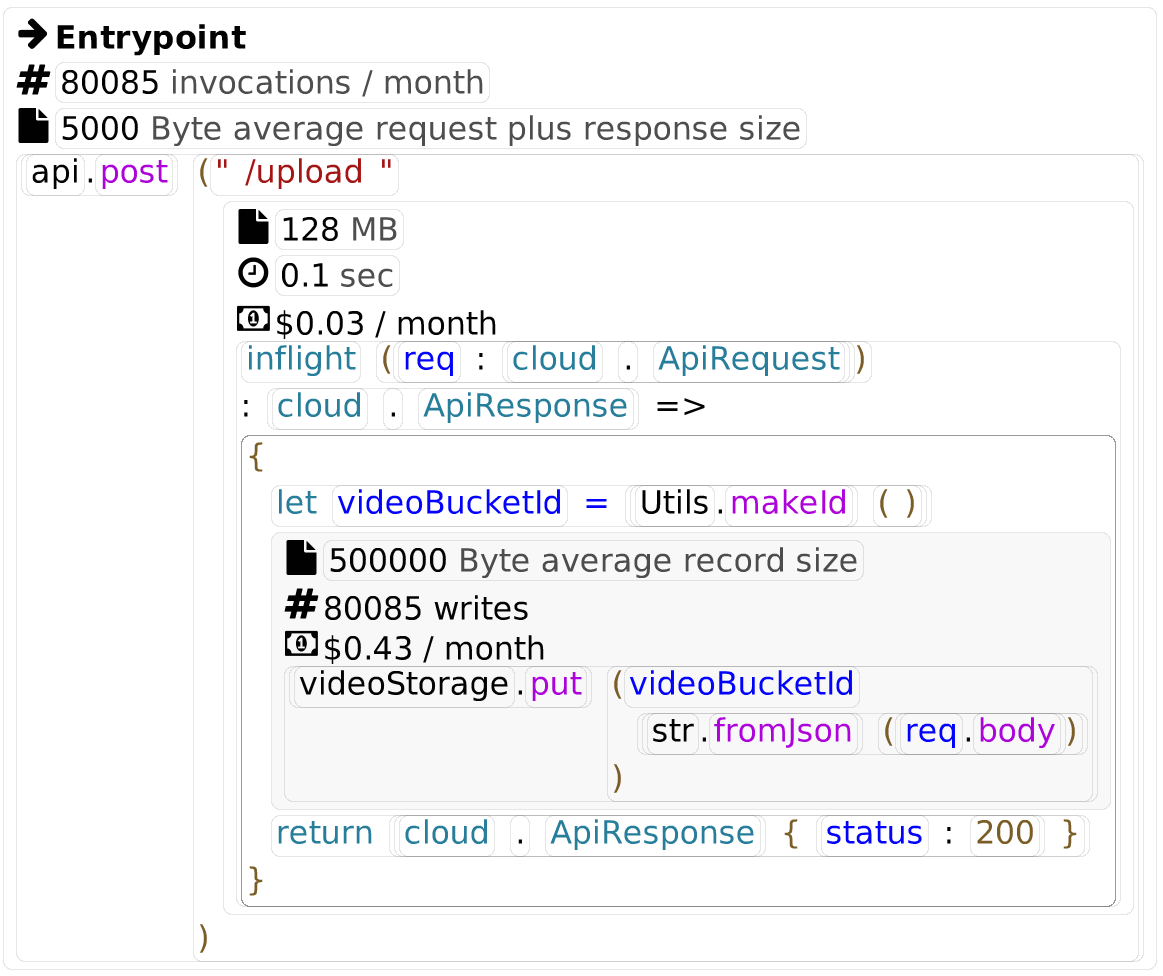}
    \caption{HTTP endpoint and handler annotated with our proposed user interface for communicating costs.} 
    \label{fig:ui-entrypoint}
\end{figure}

% 1 establish the importance of this research topic
Cloud computing accelerates application development and deployment by providing developers with rapid infrastructure provisioning and managed services \cite{avramAdvantagesChallengesAdopting2014}.
Instead of owning physical servers, developers can opt to lease infrastructure and services on-demand, allowing them to compose their applications as a mixture of self-written and managed services and pay only for what resources they actually use.
Case studies have reported that resulting cost savings for hosting can reach 77\% and up to 95\%, depending on the given cloud application \cite{villamizarCostComparisonRunning2017, adzicServerlessComputingEconomic2017}.

% 2 general background information
Cloud infrastructure and services come with various pricing models, ranging from flat monthly costs (subscription-based billing) to pay-per-use models, each having many cost-related configuration options (usage-based billing).
While such fine-grained models can offer economic advantages, they also introduce opaqueness regarding the origin of costs, increasing the risk of unplanned expenses \cite{eivyBeWaryEconomics2017}.

% 3 same as 1 and 2 but in more detail
% Example of failed costs
For instance, a simple modification in a cloud application's code could significantly increase expenses if it triggers multiple calls to costly services like a secret manager or unexpectedly triggers a chain of computationally intensive services.\footnote{Accounts from industry: https://news.ycombinator.com/item?id=31907374 https://twitter.com/donkersgood/status/1635244161778737152}
% Examples:
% https://news.ycombinator.com/item?id=31907374
% https://twitter.com/donkersgood/status/1635244161778737152
% https://shopify.engineering/reducing-bigquery-costs
% A developer working on a cloud application has to consider cost information alongside traditional resources to develop cost-efficient applications and avoid unexpected costs \cite{eivyBeWaryEconomics2017}.
Since the motivation for opting to move services to the cloud often arises from a need to support large amounts of traffic, designing cost-efficient applications is an important concern for developers~\cite{eivyBeWaryEconomics2017}, as even small architectural decisions can have a large impact when called millions of times.

% 4 general problem area or the current research focus of the field
Currently, obtaining cost information for an application is usually done in two ways:
First, before the services and resources are provisioned, developers can use web-based cost calculators, which require error-prone manual mapping, detached from the actual development process.
Second, after the application has been deployed, developers wait up to 24 hours for cost reports from their cloud provider, introducing long feedback loops and requiring developers to analyze the report retrospectively in a web portal, again detached from the actual development process.
% And some vendors also provide anomaly detection services.
% 5 transition between the general problem area and the literature review
Both options do not allow developers to efficiently understand and optimize costs of their cloud applications, especially when considering architectural changes.

%Because of that, research has been conducted to improve expense feedback cycles and cost transparency of cloud applications.
% 6 brief overview of the key research projects in this area
%For example, \citet{kuhlenkampCostradamusCostTracingSystem2017} uses a log-based approach to analyze cost information of cloud applications retroactively.
%Whereas, \citet{leitnerModellingManagingDeployment2016} proposed a cost model and interface integration, however, with the need to manually create and update the model.
% 7 research gap description
%Despite the improvements, developers continue to rely on source code detached cost models or post hoc analysis to understand where costs are coming from in their code.

Approaches in research suggested using logs for a more fine-grained post-hoc analysis~\cite{kuhlenkampCostradamusCostTracingSystem2017} or even proposed a model that would allow obtaining a cost prediction based on parameters provided by the developer~\cite{leitnerMixedmethodEmpiricalStudy2019}.
While supporting developers in comprehending expenses, both approaches detach insights from source code, forcing developers to bridge the gap between code and associated costs themselves.

% 8 describe the paper itself
To close this gap, we present an approach to automatically derive a cost model from application code based on static analysis of a cloud application following the Infrastructure-from-Code (IfC) paradigm\footnote{https://klo.dev/state-of-infrastructure-from-code-2023/}, where infrastructure declarations are derived from source code. %\todo{Add reference for IfC, even if it's just some online article. Then maybe as a footnote instead of a reference}
Building on that model, we present a user interface that displays derived cost information attached to the source code expression causing the expenses, with the goal of enhancing cost comprehension on a per-expression basis (see \autoref{fig:ui-entrypoint}).
% 9 details about the methodology reported in the paper
% 10 announces the findings
% 11 What does it mean 
The contributions of this paper are thus: (1) increasing cost-transparency of cloud applications, (2) supporting developers in making informed cost-based architecture decisions, (3) reducing the risk of unexpected costs while having a short feedback loop without the burden of manually deriving a cost model, and (4) laying out the foundation for cross-vendor cloud application cost models.
The prototype we describe in this paper is open-source and available on Github\footnote{https://github.com/hpi-swa-lab/sb-tree-sitter/tree/master /packages/Sandblocks-Wing}.

% Paper structure
%\todo{LA: Diesen Paragraphen können wir ggf. streichen, falls wir den Platz benötigen}
%In the following, we provide background information about cloud applications, their pricing models, and deployment options.
%We describe a cost model for calculating expenses related to a given cloud application.
%Using the cost model, we describe how we drive the model parameters from an application using a static analysis approach.
%Additionally, we discuss how we can integrate the cost information into a user interface.
%Finally, we discuss the applicability and limitations of our model and user interface integration, future work, and conclude the paper.

\section{Background}

In this section, we will briefly describe the domain of cloud computing, pricing models for cloud computing, followed by approaches to deploy cloud applications.

\subsection{Cloud Computing}
\label{subsec:cloud_computing}
% General description
Cloud computing describes an on-demand delivery model of computational resources over the internet.
Instead of providing and maintaining its own infrastructure to run applications, developers provision servers and services from cloud providers.
% IaaS, PaaS, SaaS
% Cloud services are categorized as Infrastructure-As-A-Service (IaaS), Platform-As-A-Service (PaaS) or Software-As-A-Service (SaaS) based on the control the provider and developer have.
% Thus, IaaS applicaitons such as the provisioning of physical services allow developers to use an entire server, however they are also responsible for maintenance of the operating system and required runtimes.
% PaaS ...
% SaaS ...
% Compute vs. Storage
% A second way of categorization of cloud services is the seperation of services for \emph{compute} and \emph{storage}.
% Compute service include the provisioning physical services, virtual machines and serverless functions.
% Compute
% - EC2
% - VM
% - Lambda
% Storage
% - S3
% - Databases
% - Caches
%Scalability and Elasticity
In addition to a large portfolio of services, cloud computing is especially interesting for developers because of provided \emph{scalability} and \emph{elasticity} of many cloud services.
\citet{lehrigScalabilityElasticityEfficiency2015} describes both terms as follows:
Scalability refers to the ability of a service to automatically adjust its resources to increase its maximum processable workload without compromising on service quality.
On the other hand, elasticity describes how well services can adapt their maximum processable workload over time.

% Serverless functions as example for scalable/ elastic services
% TODO emphasize the general challenge of cloud, using the example of serverless (loosely connected, distributed)
The most prominent on-demand paradigm in the cloud leveraging scalability and elasticity are serverless functions.
Serverless computing is a cloud computing paradigm that allows developers to create applications without the need to manage servers and scaling capacity.
% More detail what a developer is doing
The main building blocks of serverless applications are so-called cloud functions, also known as Function-as-a-Service (FaaS) offerings.
Cloud functions are small, stateless programs executed based on external triggers such as HTTP requests or state changes in a database \cite{vaneykSPECCloudGroup2017}.
Developers assemble larger applications using cloud functions combined with vendor-provided services such as databases, API gateways, or queues.
Cloud functions are dynamically scaled to meet workload demands by increasing or decreasing available resources, ensuring optimal resource allocation for the given workload.
Because of their built-in scalability, cost estimation and monitoring of serverless applications are particularly important.
%\todo{elasticity wird nicht definiert; gute Paper mit Definitionen: https://dl.acm.org/doi/10.1145/2737182.2737185}

\subsection{Pricing Models}

One advantage of building applications following the serverless paradigm is that cloud functions do not incur any cost as long as they are not being executed.
The underlying pricing model is called \emph{usage-based billing}, where the cloud provider only charges for resources during their usage.
%, e.g., by the millisecond, based on a base price determined by the compute power available to the application.
Synonyms for usage-based billing include consumption-based billing, pay-per-use, and pay-as-you-go.

\emph{Subscription-based billing}, on the other hand, means that one pays a recurring fee for a fixed period of time, granting access to a specific configuration such as a virtual machine with a certain number of CPUs, a certain amount of main memory, etc.
That fee is a flat rate regardless of resource usage on the provisioned virtual machine.
A synonym for those virtual machines is ``reserved instances''.
Discounts are available for longer commitments, e.g., one to three years.

Besides these two general billing approaches, there are also hybrid settings with a fixed monthly rate plus usage-based components or special offers such as free tiers.
Finally, cloud providers offer unused capacity at a discount via so-called transient or spot instances.
Those discounted resources can, however, be reclaimed if the provider needs the capacity due to increased demand.

% Pricing
%Each of the services possesses a tailored pricing model.
As mentioned above, serverless offerings usually come with usage-based billing.
A common approach is that cloud function prices are determined by the execution time measured in milliseconds multiplied by a base price resulting from the configured main memory.
The configured main memory then again determines the available CPU power for the application.
Serverless storage offerings such as document databases might bill the developer by the number of read and write operations, the number and size of stored documents, and the specified quality of service.
While read and write operations can be scaled down to zero, once data is persisted in the database, there will be a recurring fee regardless of usage.
This indicates that understanding usage-based billing is complex and involves considering many different cost factors.
% Advantage and Disadvantage of Pricing Model
An advantage of this pricing model is the auto-scaling mechanism.
If no workload is required, the application does not cost anything (at least in theory, see the database example above).
If more computing or storage power is required, the resources can easily be scaled up.
This pricing model does, however, also introduce cost in-transparency due to the large amount of additional cloud provider services that all have their own pricing models, such as queues, configurable databases, and cloud functions. 
%\todo{Add some examples for services, e.g., based on the diagram I saw before using the AWS pictograms}
The number of possible configurations, the difficulty in predicting future workloads, and the fact that many ``metered'' services are usually part of serverless applications, are all factors making the development of cost-efficient serverless applications a challenging endeavor.

\subsection{Methods of Cloud Application Deployment}
\label{sec:methods_of_cloud_deployment}
% Problem: Services need to be provisioned. But how?
In a cloud environment, infrastructure and consumed services must be provisioned before usage.
% One service
For simple use cases, such as deploying a single web service, a developer can manually provision the service using the vendor's online dashboard.
% Problem and transition to Declarative Files
This may quickly become difficult to oversee and reproduce for sophisticated applications using multiple deployments that may span dozens of individual services and complicated permission structures.
% Transition to declarative and imperative appraoches
In response, declarative and imperative approaches for provisioning cloud infrastructure and services arose.
Declarative automation systems such as Ansible for single machines or Terraform for clusters of machines allow developers to state the desired state in a domain-specific language (DSL) declaratively.
Tools interpret the declarative description of infrastructure to provision cloud environments automatically.
Automation systems such as AWS's Cloud Development Kit (CDK) enable developers to use an imperative approach instead of declarative DSLs.
This allows the developer to use pre-constructed abstractions in a given programming language, such as an object representing an object store to declare the required infrastructure.

% Transition to Wing as integration in code
Both approaches introduce redundancy: a separate declaration of services duplicates the same abstract services, as used in the source code, from which parts of their declaration could be inferred.
Notably, aspects such as host, port, and schemas for a database are commonly provided in the application's source code.
Other aspects related to source code, such as memory requirements for services like cloud functions, are not redundant but detached from source code in configuration panels or separate declaration files.

% Wing
To bring the configuration of the deployment closer to the source code that concern it, IfC approaches are being developed to merge infrastructure and service declaration into programming languages to reduce this redundancy.
%\todo{MA: Dieser objektorientierte Ansatz ist aber nur eine mögliche Lösung für IfC, oder? Ich verstehe die folgenden beiden Sätze auch nicht richtig.}
An IfC approach detects explicit uses of services and infrastructure from source code, which in non-IfC code are mere proxies communicating to an endpoint, and derives a declaration for deployment.
For other deployment concerns, which are not usually expressed in code, IfC approaches use annotations or similar means of adding configuration to code.
% The object acts as a declaration that an automation system can use to prepare the required infrastructure.

An example of an IfC programming language is Wing \cite{monadainc.WingDocumentation2023}.
\autoref{lst:pop} shows an example of this work-in-progress programming language, which was developed to ease the development of cloud applications \cite{monadainc.WingDocumentation2023}. 
The compiler of Wing finds instantiations that require allocation of cloud services and automatically derives a description for Terraform or AWS CDK, allowing the developer to focus on the source code instead of managing and understanding separate infrastructure declaration files.

\begin{comment}
%\begin{tcblisting}{breakable,listing only, listing options={language=Wing}, size=fbox, colback=lightgray}
\begin{lstlisting}[language=Wing, caption={Wing example in which a queue and a bucket for object storage are instantiated in an object-oriented manner (adapted from \cite{monadainc.WingDocumentation2023}).},label=lst:winglang-example]
bring cloud;

let queue = new cloud.Queue(timeout: 2m);
let bucket = new cloud.Bucket();

queue.addConsumer(inflight (body: str): str => {
  let key = "myfile.txt";
  bucket.put(key, body);
});
\end{lstlisting}
% \end{tcblisting}
\end{comment}

As IfC approaches bring code and infrastructure concerns close together, they facilitate a fine-grained understanding of resource allocation and use of those resources.
Additionally, as allocation, configuration, and use of those resources occur in a single programming language, reasoning for static analysis is simplified, as only mechanisms of one language need to be supported, and the language can enforce strong references between declaration and use of resources.
For these reasons, we chose Wing as platform and language on which to build the reference implementation of our cost estimation approach.

\section{A Model For Costs for Cloud Applications}
\label{subsec:model-costs}

\begin{figure}
    \centering
    \includegraphics[width=0.9\columnwidth]{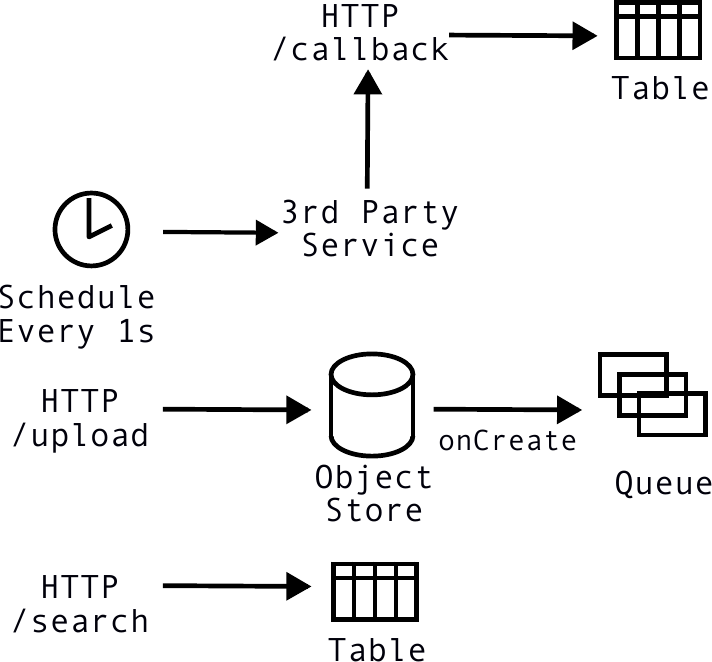}
    \caption{Illustrative schema of our running example of a video transcription service.}
    \label{fig:running-example}
\end{figure}

% Introduction of the running example
To illustrate our cost modeling approach, we will use a running example throughout the rest of the paper.
This running example is a (simplified) video transcription service with subtitle extraction, which is visualized in \autoref{fig:running-example}.

The service has two public HTTP endpoints: one that returns a download link of a video and its extracted subtitles from a database (\lstinline{/search}) and another that allows users to upload video files (\lstinline{/upload}). %\todo{Was ist ein "record for the download link"?}
The video files are saved in an object storage service.
Once saved, the service passes a public URL to a third-party transcription service.
The transcription service has a rate limit of one request per second that our application respects through a queue.
Once the transcription service has completed transcribing the video, it calls a third HTTP endpoint that receives the video identifier and the transcribed text and stores both in the database.
%Both, computational steps and resulting storage usage for the transcription service will incur costs. %\todo{Parts is unklar. Vieleicht ``all steps in this transcription process''?} 

As described in \autoref{subsec:cloud_computing}, to estimate the cost of invoking the endpoints and storing video files, we need to consider multiple factors concerning the use and configuration of our cloud application.
To facilitate this estimation process, we construct a cost model.
Our cost model is a directed graph where nodes are all application parts that incur costs.
Nodes could be a database, a queue, an API call, or an invocation of a cloud function.
Edges between the nodes in the graph correspond to the cloud application's control flow.

Nodes without incoming edges are called entry points. 
Entry points are invoked either by external users or by time-based triggers.
A typical example of an entry point is an API gateway, which forwards an HTTP request to a cloud function that is another node, creating a directed edge between both nodes.

Each node has associated costs set by the respective cloud vendor, which we define as \emph{cost factors}.
A cost factor is a single atomic aspect that influences the overall costs of the given node.
%\todo{MA: Passive voice im nächsten Satz macht es unklar, we diese Faktoren hinzufügt und wer die Knoten besucht}
For instance, a typical cloud function has two cost factors: execution time and memory usage.
A cost factor is associated with a concrete pricing model defined by a cloud vendor.
To instantiate our cost model according to the pricing model of a specific vendor, a developer has to define a mapping between types of nodes and the pricing model once.
This mapping can then be reused for instantiations of cost models for the same vendor.
% The association of cost factors of available services from cloud vendors to the IfA is a one-time effort.
%A cost factor is added to our total operational cost each time we visit the node when tracing an invocation along the control flow that the graph models.

We identified three categories of cost factors: invocation, fixed, and accumulating cost factors.

\begin{description}
    \item[Invocation cost factors] incur cost each time its associated node is reached during an execution ("invoked").
    Examples of invocation cost factors are execution time and required memory for cloud functions or byte size of a data transfer.
    Invocation cost factors are associated with usage-based billing offerings of cloud providers.
    % This includes most serverless offerings of cloud vendors.
    \item[Fixed cost factors] incur cost regardless of use and are solely related to the provision of a service, such as a physical server or database, which is charged at a fixed rate, as in subscription-based offerings.
    \item[Accumulating cost factors] change over time based on the use of a service.
    For example, the expenses associated with data storage increase as an application writes data to the database.
    Accumulating cost factors occur when using usage-based billing offerings for storage solutions.
\end{description}

%Cost factors are numeric values or configuration parameters.
While some factors are directly determined by the pricing structure of a vendor, such as the cost per invocation of an API, and some are evident from code, for example, whether a database write occurs within a transaction or not, others need to be estimated or looked up by the user or automatically retrieved from analytical data, such as visitor numbers.
Based on this observation, we further distinguish between \emph{external} and \emph{internal} cost factors.

\begin{description}
    \item[External cost factors] are those determined by actors external to the system.
    For example, the number of requests to an HTTP endpoint might depend on visitor numbers or the payload size on the length of videos uploaded to our system.
    Developers may make guesses to allow the model to perform calculations or consult historical data.
    % \item[External Factors] are those that the system cannot derive by itself.
    % For example, the number of times the homepage of a website is opened per day, or the average length of text that users enter into the comments form.
    % To assess these values, the developer may refer to historical monitoring data, predictions, or make a guess.
    \item[Internal cost factors] are all other factors and can typically be inferred or automatically estimated.
    For example, the duration of a cloud function can be estimated by running it with representative payloads.
    Or, the storage taken up by a database can be inferred through the number and parameters of insert calls that the code will make.
\end{description}

%The concrete factors for a given part and their numeric cost will depend on the vendor's pricing structure we are choosing, e.g. whether we have a reserved term or pay per use.
%Consequently, our graph model is defined independently of a concrete pricing structure but can only be instantiated once we have settled for a pricing structure we want to analyze our application for.

External cost factors appear at entry points, where user interactions with the cloud application occur.
Internal cost factors commonly depend on external cost factors for their calculation.
For example, if profiling the runtime duration of our example video transcription service with short ten-second video snippets, the calculations will differ if users upload videos that are hours long.

\begin{figure}
    \centering
    \includegraphics[width=0.9\columnwidth]{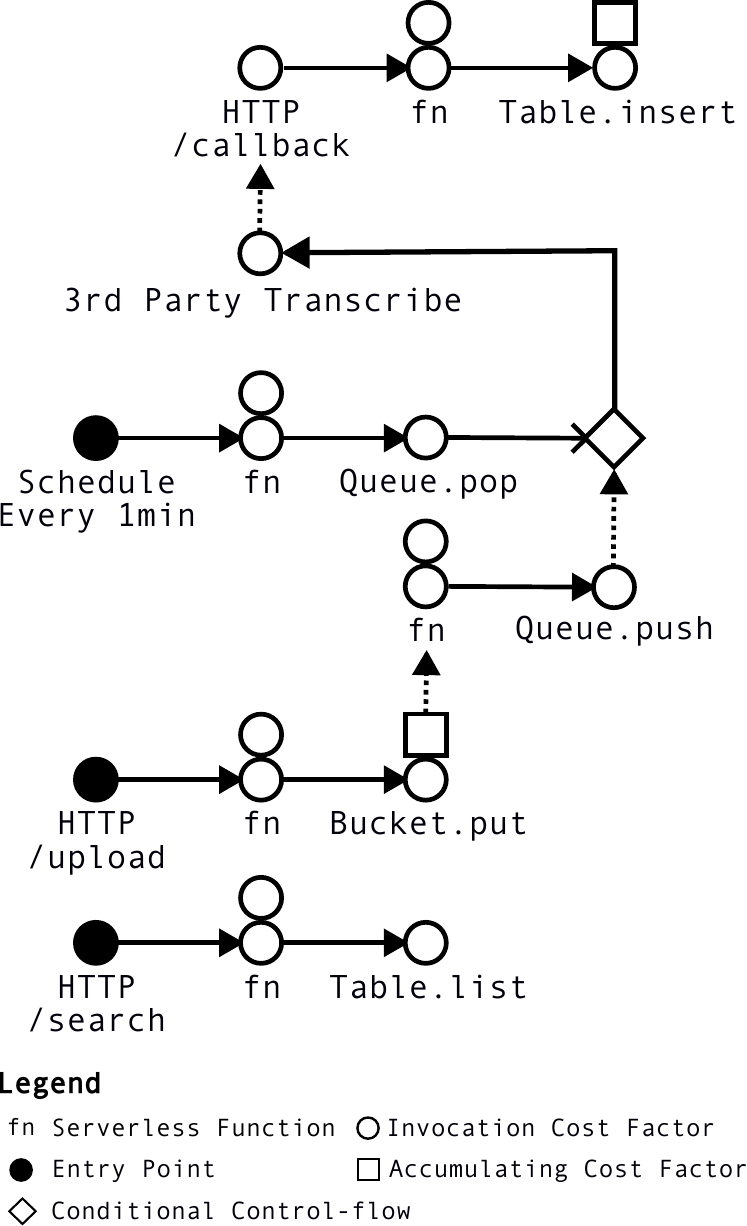}
    \caption{Cost model for our example video transcription service.
As syntax, we propose to use circles for invocation factors and squares for accumulating factors.
Filled arrows denote synchronous control flow.
Dashed arrows denote a jump but continuation of logical control flow, such as through a deferred trigger.
A conditional control flow element has a filled and an open arrow. The filled arrow denotes the dominant flow, which should be considered for the invocation count of subsequent nodes.
The open arrow denotes the synchronous control flow that allows us to reach the diamond but may be invoked at a different frequency.}
    \label{fig:model}
\end{figure}

% Cost model explaination
The cost model for our running example is shown in \autoref{fig:model}.
Each service incurring costs is represented as a node with the associated cost factors.
In our example, most cost-incurring nodes have invocation cost factors.
Serverless functions (fn) have two invocation cost factors: allocated memory and runtime, whereas method calls \lstinline{Table.list}, \lstinline{Queue.pop} and \lstinline{Queue.push}, as well as HTTP endpoints (\lstinline{/upload /search /callback}) have costs per execution.
Nodes for calls to storage solutions such as \lstinline{Bucket.put} and \lstinline{Table.insert} are associated with both invocation and accumulating cost factors since they are billed per invocation, and the resulting storage entry accumulates costs over time.

To estimate the costs for a single execution of an entry point, we simulate the expected control flow a program takes.
Therefore, we traverse the cost model graph beginning from the entry point and sum costs for each cost factor of each passed node. 
All cost factors are combined to calculate the costs for one node using the rules determined by the vendor's pricing model.
This can range from a simple multiplication of factors to more complex calculations that increase in steps as thresholds of use are exceeded.
To assess the total cost of running the complete system over a given time frame, external cost factors determine how often the entry points of our application are invoked and, thus, how often we need to simulate execution and add up the resulting cost.

Some edges in the graph are conditionally taken based on the occurrence of certain API calls elsewhere rather than their control flow as expressed in code.
We call these implicit control flow edges.
Implicit control flow edges occur when a relationship between the source of control flow and the target of control flow is not stated explicitly in the source code.
For example, debounced processing of work through a queue establishes an implicit connection between the place where work is pushed to the queue and the place where work is popped from the queue.
In the visualization of the model, we add two arrows for such scenarios that meet in a diamond, as seen in \autoref{fig:model}: one arrow shows the actual control flow that triggers the target of the implicit control flow; a second arrow shows the implicit control flow from the source.
For the purposes of our cost model, the first arrow showing the actual control flow is irrelevant and only helps the developer understand the relationship.
For example, the frequency of popping values off of a queue is considered secondary to the frequency of pushing to that queue when considering a branch that is only taken when the queue is not empty.

%\todo{HA: Macht es vielleicht Sinn die Cost Factors bereits oben in der Section zu den Pricing Models einzuführen? - Da das unsere einteilung ist und keine von jemand anderen macht es wahrscheinlich mehr sinn die hier aufzuführen oder?}

%\todo{MA: Wo kommen die executions her? Instrumentieren wir Winglang dafür? -> wird unten erwähnt, aber sollte hier kurz zusammengefasst werden}

To summarize, we mapped our running example as a concrete cost model, as shown in \autoref{fig:model}.
Here, the entry points determining the overall traffic are the HTTP routes that allow users to upload videos or search transcriptions.
As the nodes are invoked along the execution of the model, their associated cost factors incur expenses.
First, the HTTP endpoint incurs expenses for its invocation and cost for the bandwidth for request and response.
Next, the HTTP endpoint invokes a function to handle the request, which incurs two cost factors for memory and execution duration.
Following, the function places the video file in a bucket store, incurring a cost factor for the upload and another accumulating factor for the data added to the store.
The synchronous flow of the function ends now, but our logical flow continues through a trigger reacting to files being uploaded to the bucket by pushing to a queue.
Now, the second entry point is of relevance, as it will attempt to pop an item off the queue every minute and, if an item had previously been pushed, send that file to a third-party transcription service.
That service sends a completed transcript via an HTTP request to our infrastructure, where we store the transcript in a database table.

% How do we calculate costs?
% We calculate the costs associated with the involved nodes and also account for the costs incurred from storage.
% Total costs
\section{A User Interface for Visualizing Cost of Cloud Applications}

As described in \autoref{sec:intro}, our goal is to visualize the cost information of a cloud application as close as possible to the code causing the costs.
To derive the cost model of a cloud application introduced in \autoref{subsec:model-costs}, we first need to derive the graph structure from the source code (see \autoref{sec:extract-model-from-code}).
We then use the resulting model to construct a user interface to visualize costs directly within the source code (see \autoref{sec:user-interface}).
\autoref{fig:ui-queue} shows an example of the user interface we propose.
To facilitate tight integration between the user interface and the source code, we use the Sandblocks structured editor~\cite{sandblocks}.
While using a structured editor for our approach is not required, it simplifies the creation of a user interface next to the source code.
%Each language construct in the code representing nodes in the cost model is wrapped by a decoration that communicates aspects relevant to its cost.

\subsection{Extracting the Cost Model from Code}
\label{sec:extract-model-from-code}

In the following, we describe how we extract the cost model from Wing source code.
As discussed in \autoref{sec:methods_of_cloud_deployment} IfC approaches like Wing combine application code and declarative infrastructure construction in one source code file.
% Wing: Preflight vs. Inflight 
To facilitate this, Wing distinguishes between a \emph{preflight} and an \emph{inflight} phase, as seen in \autoref{lst:pop}.
By default, code runs in the preflight phase.
Code running in the preflight phase evaluates and obtains the infrastructure's declarative descriptions, such as the instantiation of an API gateway or a database.
Code marked with inflight is contained in closures that are executed at runtime on a provisioned cloud server.

Preflight API calls in Wing wire together different parts of the infrastructure to form its declarative description that can be deployed.
In particular, preflight code schedules when inflight code will be run.
For example, a preflight call may set an inflight function to run every time an object is added to an object store.
Furthermore, preflight API calls declare the use of resources, such as a static host of files for the web or a database that charges for the amount of storage taken up.
Inflight code uses resources that are declared in the preflight phase: an inflight closure might reference a database and insert a record into it.
To extract a cost model from Wing code, we follow a process of three phases:
\begin{enumerate}
  \item we find constructors of resources in the code by looking for specific AST nodes (e.g., \lstinline{new cloud.Api()}),
  \item we find all calls on instances of these resources by statically analyzing the use of the variables that the resources are bound to, and
  \item we reference a list of parameters relevant to cost estimation for each resource and call on that resource.
\end{enumerate}
Once we collected the set of resources and API calls, representing nodes in the cost model graph, we derive the control flow between the nodes through a list of hard-coded rules derived from the function of the Wing APIs.

The control flow begins from an entry point, linearly executes code in an inflight closure, and exits, except for two cases.
First, event-driven triggers might leave the linear control flow as manifested in sequential code.
For example, we know that an upload to an object storage will invoke all handlers that are subscribed to the corresponding upload trigger:
\begin{lstlisting}[language=Wing, caption={HTTP endpoint adding a file to a bucket and a trigger responding to files added.},label=lst:upload-push]
api.post("/upload",
  inflight(req: ApiRequest): ApiResponse => {
    videoStorage.put(str.fromJson(req.body));
    return ApiResponse { status:200 };
});
videoStorage.onCreate(inflight(key: str) => {
    queue.push(key);
});
\end{lstlisting}

% The second exception, as described in \autoref{subsec:model-costs}, is related to the execution of \emph{conditional edges}.
% They might depend on a second factor that is likely more important for our purpose of estimating costs than the linear control flow. \todo{MA: Beispiele für einen solchen Second Factor? Satz ist etwas unklar. Das Beispiel ist nett, aber die Exception muss IMHO self-contained ohne die Notwendigkeit eines Beispiels beschrieben werden}
Second, implicit control flow edges may depend on an additional factor, as seen in the code below.
While the closure is invoked every second, the API call for transcription is only invoked if an element was pushed to the queue in \autoref{lst:upload-push}.

\begin{lstlisting}[label=lst:pop, language=Wing, caption={Scheduled operation checking a queue and calling an HTTP endpoint.}]
let schedule = new Schedule(ScheduleProps {
    rate: std.Duration.fromSeconds(1)
});
schedule.onTick(inflight () => {
	if let key = queue.pop() {
		httpPost("http://example.com/transcribe", {
            videoId: key
        });
	}
});
\end{lstlisting}

% \todo{Das wirkt etwas out of place? tobe: agreed, das haben wir in der discussion}
% Depending on the implementation, every branch in the control flow might invoke different nodes, thus, different cost factors.
% For instance, if our application had a cache, our simple static analysis would disregard the branching point checking for whether the cache is hit and place all nodes of both branches in a linear sequence in the model.
% As part of this proof-of-concept, we focused on an analysis disregarding branches, except for the conditional edge scenario above.

\subsection{User Interface}
\label{sec:user-interface}

The cost model serves the vital function of calculating expenses linked to the cost factors of a cloud application.
This process is essential to improve awareness of the expenses of cloud applications.
A user interface that tightly integrates with the source code is then responsible for communicating to the developer: 

\begin{enumerate}
    \item what statements cause costs,
    \item what the overall summation of cost per month according to the model is, and
    \item how the (non-linear) control flow causes costs.
\end{enumerate}

\begin{figure}
    \centering
    \includegraphics[width=\columnwidth]{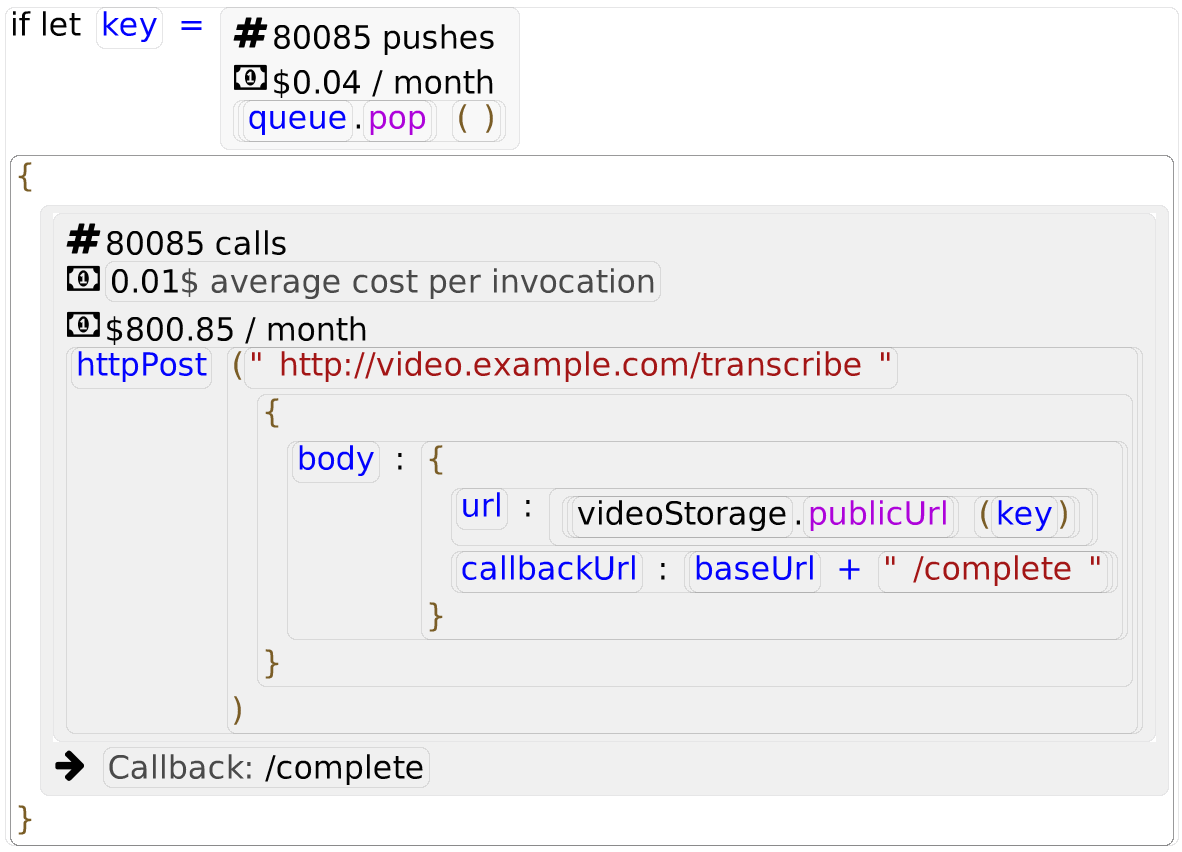}
    \caption{A screenshot of our user interface prototype. A conditional checks for elements in a queue. The API call to access the queue is annotated with the number of pushes to that queue happening elsewhere in that program and the total cost for all pop operations per month. Below, a HTTP request to a third-party video transcription service has two annotations: through the outer, the user has informed the system that this service will callback to our \lstinline{/complete} HTTP endpoint, and the other was added by the system, showing how often this call will occur, allowing the user to set a cost per call, and calculating the total cost per month. } 
    \label{fig:ui-queue}
\end{figure}

To communicate what statements cause cost, we wrap all statements and resources found during the analysis process in widgets that expose the factors of that statement that influence cost according to our model, as seen in \autoref{fig:ui-queue}.
For a database, we may show how much storage is added per month, and for an API call how frequently it is called.

\begin{figure}
    \centering
    \includegraphics[width=0.9\columnwidth]{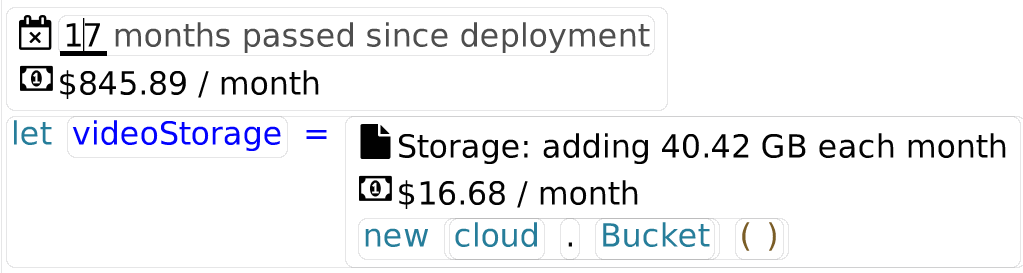}
    \caption{User interface embedded at the top of the source file. It allows users to select a month to show a prediction for and, below, how much the model indicates the monthly cost will amount to for the configured month. Below the declaration of a bucket for storing video files, showing the total of storage added each month based on all upload operations in the program and their configured payload sizes, as well as the cost for the total storage based on the point in time configured in the widget just above. } 
    \label{fig:ui-total}
\end{figure}

%\paragraph{Time}
The cost model is independent of time because it describes costs produced from a single invocation along any of its entry points.
However, accumulating factors like database storage as described in \autoref{subsec:model-costs} will often increase cost over time.
Consequently, we provide a slider as part of the user interface that allows developers to globally configure a point in time to know the cost for.
All cost factors that depend on how long the system has been in use should refer to that slider.
In our running example, the slider influences storage costs for the object store and the table.
As time progresses, entry points are triggered, leading to new entries in the object storage and table.
Consequently, accumulating costs increase over time since data entries are only added, not deleted.

%\paragraph{Total estimate for the application}
An overall cost estimate for the entire application is given at the top of the source file, next to the selection for point in time, as shown in \autoref{fig:ui-total}.
The widget adds up all costs for the configured month.
Developers will need to make sure to have sensible estimates for external factors, for example, for the amount of traffic each endpoint receives, for this total to be useful.

%\paragraph{Persistence}
All language constructs that require annotations persist the information provided by developers in the source code.
Specifically, as Wing does not have an expression-level annotation syntax construct, we wrap each invocation in an array of the form \lstinline![new cloud.Table(), {averageRecordSize: 200}][0]! that embeds the relevant information in the AST but does not affect runtime.

While we provide hard-coded rules describing the control flow of Wing's API and triggers for use by our static analysis, a call to our cloud application that occurs from a system that acts as a black box cannot be automatically inferred.
Instead, we allow developers to annotate that an expression will invoke an HTTP endpoint in the user interface, as shown in \autoref{fig:ui-queue}.
Once annotated, our system can infer the number of invocations in the HTTP endpoint based on the number of invocations to the black box service.

The user interface thus uses the cost model information to provide developers with relevant and acute information.
Changing values, either in the code or in a configurable cost factor, shows immediate changes in all derived cost factors.
A large number of relevant factors are automatically derived by our system, notably non-linear control flow, and the developer is able to fill in the gaps to arrive at useful information, even in the early stages of the presented prototype.

\section{Related Work}

Most research in the field of cost transparency for cloud applications was published in the cloud computing community. 
For instance, \citet{leitnerModellingManagingDeployment2016} introduce CostHat, a model to calculate the cost of microservice-based applications, which integrates cost information in an IDE.
While this research has similar aims, our approach differs in two key aspects.
% Automatic generation of the cost model 
First, CostHat explicitly states that the instantiation of their model is out of scope, requiring developers to create the cost model by themselves.
Our approach uses static analysis, supported by the IfC deployment approach, to derive cost models from source code.
% More detailed 
Second, CostHat simplifies the cost calculation to four factors: compute costs, costs of API calls, costs of IO operations such as writes to a database, and additional costs.
In practice, cloud vendors use a variety of factors that go beyond these four.
For example, the cost of cloud functions depends on configured memory size.
Moreover, the cost of storage and networking depends on the size of objects accessed or transferred.
Our model allows specifying arbitrary parameters, thus letting developers to match the current and future billing practices.
%Third, since our model is tightly coupled to the source code, we can provide line-based feedback and widgets to interactively explore cost directly where cost occurs in the code editor.

Another tool aiming to improve cost comprehension for cloud applications is Costradamus.
Costradamus, introduced by \citet{kuhlenkampCostradamusCostTracingSystem2017} is a tracing system focusing on retroactively capturing costs in a cloud environment.
It enriches log statements with cost-related metadata and analyzes this data to gain insights into the sources of costs.
While Costradamus dynamically captures fine-granular cost information, it has limitations in reliably capturing all possible paths of an application, as it relies on these paths being used while capturing is active.
Moreover, the approach does not allow theoretical scenario analysis and introduces long feedback loops since tracing data must be captured first.

% Transition
In addition to specific tools, other research areas are also related to our topic.
% Performance 
For instance, the integration of performance information in code editors shares a close connection with the objective of enhancing cost comprehension, as both aim to convey runtime information to users.
\citet{baltesNavigateUnderstandCommunicate2015} integrate runtime performance information in a code editor.
Based on a user study, they conclude that the integration of performance information in the IDE helped developers to find and understand performance bugs.
Moreover, \citet{citoInteractiveProductionPerformance2019} introduce PerformanceHat, which integrates profiling information from production systems into the editor code, to give direct feedback about the runtime performance of specific methods and functions.
They also identified that developers find performance bugs faster by integrating profiling information, which may be applicable to costs as well.

\citet{obetzStaticCallGraph2019} describe an approach to derive a static call graph from serverless applications based on vendor-provided SDKs.
The call graph can be used to communicate control flow in complex setups across serverless functions to, for example, assessing security or performing dead code elimination.

% Prediction
Furthermore, much research has been conducted in the context of serverless functions to predict the costs of cloud functions and workflows.
For example, \citet{eismannPredictingCostsServerless2020} efficiently predict the costs of serverless workflows using a monte-carlo simulation and information from prior executions.
However, they treat each function as a black box, ignoring possible cost side effects such as calling other services or using external data storage.

% Non-peer reviewed topics
Finally, the open-source community also published tools to ease the development of cloud applications related to costs.
\emph{Infracost} \cite{InfracostInfracost2023} is a specialized tool to estimate costs for infrastructure based on a declarative Terraform file, leaving out runtime information.
Resulting cost information is either displayed directly in the code editor via plugins or is part of the version control workflow, enabling developers to discuss changes on the cost before they occur.
The tool AWS Lambda Power Tuning \cite{casalboniAWSLambdaPower2023} enables developers to experimentally execute functions to determine the configuration with an optimal cost-performance ratio.
%Increasing the memory of a cloud function results in faster execution speed due to automatic allocation of faster CPUs by cloud vendors, but at higher costs.

\section{Discussion and Future Work}

In the following section, we first discuss the characteristics of the cost model and, second, the user interface to communicate the costs derived from the model.
In addition, we identify future work for both cost models and the user interface to increase awareness of the expenses of cloud applications.

\subsection{Cost Model}
The model, as presented in the paper, acts as a proof-of-concept for further development.
Below, we outline its current strengths and limitations.

\paragraph{Flexibility}
The model can simulate execution and consider relevant factors for cost estimation of cloud applications, including statically derived control flow and logical connections, such as invocations made from third-party systems.
Our proposed directed graph approach consisting of nodes and cost factors allows instantiating models for current and future pricing models of different vendors.
Currently, a static mapping between the pricing of services for each vendor to the language constructs once.
In future versions, adapters to the vendor pricing APIs can be used to instantiate our model.
However, when having a model with up-to-date pricing information of multiple vendors, our system can automatically estimate expected expenses for each vendor, allowing the developer to make informed decisions before selecting a vendor for their cloud application.
At the same time, a challenge to our approach is the complexity and vastness of the configuration space offered by cloud vendors.
For example, current cloud developers can select quality-of-service levels, compute units, regions, and dozens of other factors, to configure a database.
Pricing tiers, which decrease costs per unit as usage increases, make it more difficult to predict the effect of changes in usage, as they no longer grow linearly.
Communicating this large configuration space effectively and concisely will be important in future work.

\paragraph{Control Flow}
To precisely calculate a cloud application's costs, identifying possible control flow branches, particularly costly rare worst-case branches, is vital for a precise estimate.
While the model incorporates basic control flow from the source code of a cloud application automatically and allows developers to patch non-inferrable gaps, the presented implementation currently does not interpret control statements like loops for if-statements.
For instance, when a cloud function is called within a loop, the current model does not reflect multiple executions, missing relevant cost information.
Similarly, a branch may let control flow make two different API calls based on user input, or a query may cache results to reduce costs by avoiding subsequent computations.
In future versions, the model should detect such cases and create conditional edges that can be weighted by probabilities in the form of the user's input or a cache hit ratio.
To derive places where probabilities are necessary and what data they depend on, a complete version of the static analysis might incorporate symbolic execution and abstract interpretation.
Both techniques might help developers understand dependencies between input and branches along the control flow and identify possible costly branches of a program without executing it.

%\paragraph{Redundancy in External Factors}
%External factors in the current implementation are constant values.
%However, the same external factor may appear in two places or benefit from being calculated as an expression, for example the configured size of a request payload and the size of record stored in a database from that payload.
%In future work, we want developers to express such relationships as part of their program, e.g., by using a domain-specific language.

% TODO attribution of fixed costs (across all uses vs separately)

\subsection{User Interface}

The user interface visualizes the expenses of a cloud application by using the cost model to increase the cost awareness of a developer.
The possibilities of interactions exposed in the user interface determine how developers can interact with the model.

\paragraph{Proximity to Source Code}
Similar to the goal of IfC to combine infrastructure declarations and source code and remove redundancies, our user interface brings cost information directly to the source code.
Through annotations, information developers add to aid cost estimation persists in source code and can be shared.
Thus, developers can experiment and understand the impact on cost interactively.
In future iterations, lessons from live programming~\cite{live-study} could be integrated to support developers in quickly experimenting with permutations of their program.

\paragraph{Manual Predictions}
In the current user interface, many parameters can be derived.
Consequently, the burden for developers when estimating cost, which as described either required developers to enter every detail in cost calculators or wait for billing information from deployments to arrive, is significantly lessened.
As the prototype has an understanding of control flow, it can propagate factors that are configured once to multiple places where they are used.
Still, some manual data entry is required to calculate costs that may not be necessarily required.
Recent studies show that some of the parameters of serverless applications, such as the execution time of a serverless function, can be predicted accurately \cite{eismannPredictingCostsServerless2020}.
Similarly, we can include production information in the estimation analogous to performance research \cite{baltesNavigateUnderstandCommunicate2015}.
For example, tests or examples such as in Babylonian Programming~\cite{babylonian} could be used to perform trial runs of functions to estimate duration and memory usage.
% In addition, speculative execution of functions could lead to 
% Ideally, a developer should only need to provide external parameters, and based on these, all internal parameters would be derived.
% Even with external factors, visualizations could help developers by displaying trends or histograms that communicate distributions of payload sizes or expected workload.

\paragraph{Visualization for Costs Tracing}
Currently, our user interface does not aid developers in analyzing the sources of a factor.
Although the necessary data is already available, the user interface does not yet communicate the edges between the model nodes.
For example, the user interface may display the edges overlayed on the code or show a tabular view.
% It may also not be straightforward to display complicated dependencies in a concise, understandable manner.
Visualizations attached to the code causing costs might enable developers to explore the cost factors of their application interactively.
More comprehensive analytics could highlight where cost optimizations may have the most impact.
This expanded approach would empower developers with more detailed information and analysis options.

%\paragraph{Programmatic Cost Calculations}
%In addition to visualizations, a developer might want to program relationships between parameters influencing the cost factors of an application.
%For example, reusing a variable for the average size of a video decreases the amount of redundant configurable parameters.

\paragraph{Optimizing Cost}
Given a complete model, a future direction could begin identifying parameters to optimize for cost efficiency.
% We imagine that based on cost estimates, a compiler for cloud applications could consider cost estimates and make adjustments to the application or to cost-related configurations to minimize expenses.  
For example, developers could formulate service-level requirements such as response times and leave parameters of cost factors free to be specified by the system, as long as those requirements are fulfilled \cite{hellersteinServerlessComputingOne2018}.
Or, a linting mechanism on the graph could communicate the potential for batch operations to developers.

\section{Conclusion}
In this paper, we presented an approach to model cost for cloud applications independent of a specific vendor.
The resulting cost model facilitates the calculation of the overall cost of deploying and running a cloud application.
In a prototype, we demonstrate that it is possible to derive a large part of factors relevant for cost through static analysis and map the remaining factors into an interface where the developer can fill the gaps.

Our prototype displays the derived and configured cost factors attached to their related expressions in code.
Consequently, our approach supports developers in understanding how their code incurs cost and allows them to experiment with alternative design choices in code while directly seeing the impact on cost.
Our system demonstrates the feasibility of cost as a tightly integrated concern in development environments, paving the way for future developments where that information can be used to support developers in designing and optimizing their cloud applications.

% While the current version of the system is only a prototype, our proposed approach works towards cost transparent cloud development, expanding existing research on cost-efficient development by freeing developers from the need to manually update cost models or wait for delayed cost analytics.
% Through the tight integration into the source code that incurs the cost, our system supports developers in their understanding of cost factors and in making informed decisions for their application concerning cost.

\begin{acks}
  We gratefully acknowledge the financial support of HPI's Research School\footnote{\url{https://hpi.de/en/research/research-school.html}}.
\end{acks}

% \begin{acks}
% \end{acks}

\balance
\bibliographystyle{ACM-Reference-Format}
\bibliography{paper}

\end{document}